\documentstyle[12pt,epsfig]{article}

\newcommand{\E}{{\rm e}}
\newcommand{\D}{{\rm d}}

\begin{document}

\pagestyle{empty}

\date{}

\title{Aging in a simple model of a structural glass}
\author{ Luca Peliti$^{1,2}$ and Mauro Sellitto$^1$\\
         $^1$ Dipartimento di Scienze Fisiche and Unit\`a INFM, \\
Universit\`a ``Federico II'', Mostra d'Oltremare, Pad.~19,
I-80125 Napoli, Italy.\\
		$^2$ Associato INFM, Sezione di Napoli.}
\maketitle

\begin{abstract}
We consider a simple model of a structural glass, represented 
by a lattice gas with kinetic constraints in contact with a particle reservoir.
Quench below the glass transition is represented
by the jump of the chemical potential above a threshold.
After a quench, the density approaches the critical density---where
the diffusion coefficient of the particles vanishes---following
a power law in time. In this regime, the two-time self-correlation 
functions exhibit aging.
The behavior of the model can be understood in terms of simple
mean-field arguments.
\end{abstract}

\section{INTRODUCTION}
Although glasses are considered the paradigmatic example of long-lived
violation of thermodynamical equilibrium, the detailed nature of this
violation is still the subject of much debate \cite{Go,Go2}. Phenomenologically,
a glass is an undercooled liquid, quenched below a temperature
such that its viscosity exceeds 10$^{13}$ Poise. 
In this situation, the glass can be considered ``solid'' in 
the sense that its molecules rattle within a ``cage'' formed by their
neighbors and do not leave it, at least within experimental
time scales. However, even in the glassy state, the system does not reach
a time-independent statistical state,
but rather keeps evolving at a slower and slower pace as the time 
$t_w$ elapsed since its quench increases.
This is the origin of the striking aging effects observed in glasses
\cite{Ho,St}. 
Aging corresponds to the property that, while one-time quantities like
the average energy, volume, etc., appear to be invariant under time 
translations, two-time correlations and responses exhibit a non 
trivial dependence on both of their arguments \cite{Saclay}.
Similar properties have also been observed in spin-glasses 
\cite{agingSG,agingSG2}
and in fact appear quite frequently, provided one takes the infinite size limit
{\em before\/} taking the infinite time limit \cite{CuKu1}.
It now appears that aging can be described by a consistent and robust
phenomenology, in which several apparently unrelated properties are brought
together. On the one hand, in purely relaxational systems, the
lack of time-translation invariance (TTI) in the correlation functions
is associated with the violation of the fluctuation-dissipation theorem
(FDT) \cite{CuKu1}. This violation can be related to entropy production
\cite{CDK} and leads to the definition of an effective frequency-dependent
temperature out of equilibrium \cite{CKP}.

Aging properties have been thoroughly investigated within 
spin-glass models \cite{review} which are closely related to the 
mode-coupling theories of structural glasses \cite{Go,Go2}.
The spin-glass models proposed as a description of 
structural glasses lack, however, a transparent physical 
interpretation in terms of particles and involve a complex, random 
hamiltonian which is hard to justify as a description of a fluid.
Their justification is rather {\it a posteriori\/} in the sense that the 
phenomena they exhibit recall the behavior of structural 
glasses.
On the other hand a class of very simple kinetic models 
have been introduced to describe the slowing down of the dynamics 
\cite{Fran,FrBra}.
These models are defined by kinetic rules involving a selection of
the possible configuration changes (``moves'') and are therefore 
called models with constrained dynamics.
The kinetic rules satisfy detailed balance and are 
compatible with a Boltzmann-Gibbs equilibrium distribution involving 
a hamiltonian, usually chosen to be a trivial one.
The constraints are alone responsible for the slowing down of the dynamics 
because, near any allowed configuration, there are only few configurations 
which satisfy them. Since the hamiltonian is trivial, it is
easy to prepare the system in an equilibrium state. By
following its subsequent evolution, one can then collect data
on the equilibrium correlation functions, which exhibit TTI
by definition. In order to observe aging and FDT violation,
it is necessary to prepare the sample in an out-of equilibrium
state. One way to achieve this goal is to introduce a conjugate
field which plays the role of the temperature, and to simulate
the quench by a drastic change of this field.

In the present contribution, we consider a lattice-gas model introduced
by Kob and Andersen \cite{KA} and generalize it to allow particle
exchange with a reservoir \cite{KPS}. The kinetic constraints
prevent the particle from moving when it has too many neighbors: 
the detailed balance condition is satisfied if the particle cannot
move also if it {\em would have\/} too many neighbors after the
move. As a consequence, the evolution of the system slows down
and becomes sluggish when the particle density $\rho$ increases:
when $\rho>\rho_{\rm c}\simeq 0.88$ the self-diffusion constant
$D$ vanishes \cite{KA}.

We have allowed the system to exchange particles with a reservoir,
introducing therefore the intensive variable $\mu/k_{\rm B}T$ 
(the ratio of the chemical potential to the absolute temperature),
conjugate to the total number of particles. Since the hamiltonian
is trivial, temperature does not play any significant role and
we can set $k_{\rm B}T=1$ throughout. The equilibrium equation
of state $\rho=\rho_{\rm eq}(\mu)$ can be then trivially calculated.
There is therefore a critical value $\mu_{\rm c}$ of $\mu$ defined
by $\rho_{\rm eq}(\mu_{\rm c})=\rho_{\rm c}$. This value plays the role
of the glass temperature. In particular a quench is
represented by a jump in $\mu$ from below to above $\mu_{\rm c}$,
corresponding to a sudden compression.

Remarkably, this simple model exhibits a number of glassy properties
so far obtained only in much more complicated models \cite{KPS}.
One can perform numerical experiments analogous to smooth cooling,
quench, and hysteresis cycles: namely, one can let $\mu$
increase smoothly, perform a cycle, or jump suddenly
above $\mu_{\rm c}$. In the first
case the density $\rho$ first increases, then reaches a plateau
value that depends on the compression speed: the slower the
compression, the higher the value. The critical density $\rho_{\rm c}$
could apparently be reached in the limit of zero speed.
In the second case, the density  appears to follow a hysteresis
cycle whose area decreases as the compression speed decreases.
And finally, in the case of a quench, the density appears
to approach the critical one like a power law. In this
case, aging effects in the self-correlation function
$B(t,t')=\left<|\vec r(t)-\vec r(t')|^2\right>$ are evident.
We show here that the power-law relaxation of the density can
be interpreted as a consequence of the fact that
the diffusion coefficient vanishes at $\rho_{\rm c}$, and
that this in turns explains the aging effects in $B(t,t')$.
A simple mean-field-like theory is proposed, which could act as
a guide for further numerical investigations.

Section 2 contains the definition of the
model and the results of its simulation.
Sec.~3 contains a simple mean-field
theory which accounts for the behavior of the model under
quench. Sec.~4 contains a brief conclusion and hints for
further investigation. Some numerical results reported
in Sec.~2 had already been published in ref.~\cite{KPS}.

\section{NUMERICAL RESULTS}
We consider the kinetic lattice-gas model
studied by Kob and Andersen (KA) \cite{KA}.
The system consists of $N$ particles in a cubic lattice of side $L$, 
($V=L^{3}$) with periodic boundary conditions.
There can be at most one particle per site. Apart from
this hard core constraint there are no other static interactions
among the particles.
At each time step a particle and one of its neighbouring sites are 
chosen at random. 
The particle moves if the three following conditions are all met:
\begin{enumerate}
\item the neighbouring site is empty;
\item the particle has less than $m$ nearest neighbours;
\item the particle will have less than $m$ nearest neighbours after 
it has moved.
\end{enumerate}
The rule is symmetric in time, detailed balance 
is satisfied and the allowed configurations have the same weight
in equilibrium. 
Significant results are obtained when the threshold
$m$ is set to 4 for a square lattice in three dimensions.
With this simple definition one can straightforwardly proceed
to study the dynamical properties of the model at equilibrium.

Kob and Andersen \cite{KA} have studied this model starting from an
equilibrium configuration, which is obtained very simply by filling
in turn each site of the lattice with a probability equal to the density
$\rho$. In this way, TTI is satisfied by construction. One observes,
in particular, that the diffusion coefficient of the particles
vanishes above the critical density $\rho_{\rm c}\simeq 0.88$~,
and behaves like $|\rho-\rho_{\rm c}|^{\phi}$ for $\rho$ smaller,
but close to $\rho_{\rm c}$.

\begin{figure}[htbp]
\centering\epsfig{file=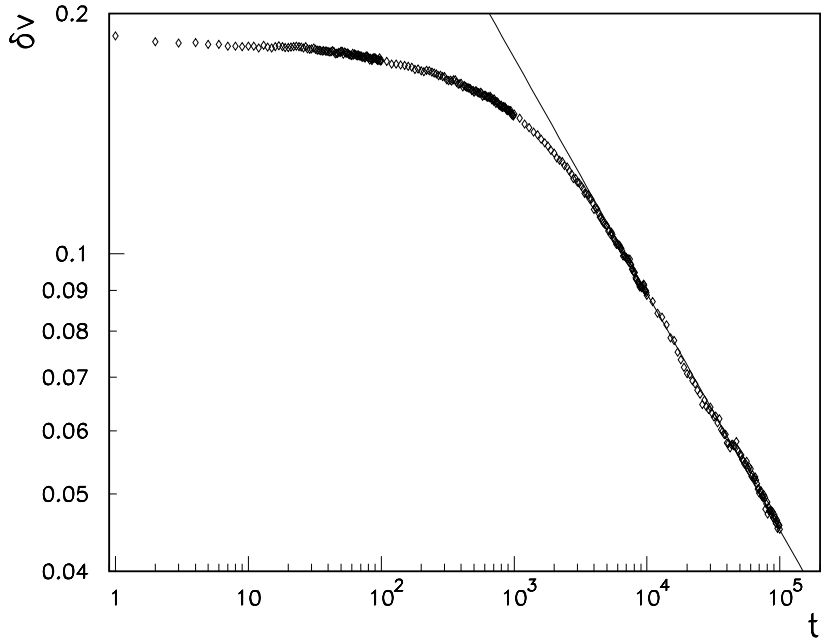,width=0.75\linewidth}  
\caption{Relaxation of the excess specific volume $\delta v=v-v_{\infty}$,
where $v=1/\rho$, $v_{\infty}=1/\rho_{\rm c}$,
after a quench to the subcritical value  $1/\mu= 1/2.2$. 
The density of initial configuration is $0.75$.
The straight line is $\delta v= 1.36 \cdot t^{-0.296}$,
with $v_{\infty}=0.88$ .}
\protect\label{relaxd}
\end{figure}

In order to study aging effects,
we have generalized \cite{KPS} the KA model by allowing the particles to
appear and disappear on a  single two
dimensional layer (the ``surface'').
This event takes place with the following Montecarlo rule:
a site on the surface is chosen at random; if it is empty we add a particle,
otherwise we remove the particle with probability $\E^{-\mu}$.
We assume $\mu>0$, since however the interesting behavior
takes place at high density ($\rho>0.5$).
Because of the periodic boundary conditions, the sample can
also be conceived as a slab, whose upper and lower surfaces
(placed, say, at $z=\pm L/2$) are in contact with the reservoir. 
For each value of $\mu$ it is trivial to evaluate the equilibrium density  
\begin{equation}
\rho_{\rm eq}(\mu)=\left(1+\E^{-\mu}\right)^{-1}.
\end{equation}
If $\rho_{\rm eq}(\mu)<\rho_{\rm c}$, the system rapidly reaches the equilibrium 
state. There is therefore a critical value, $\mu_{\rm c}$, of $\mu$, defined
by $\rho_{\rm eq}(\mu_{\rm c})=\rho_{\rm c}$. 
A quench corresponds to a jump of  $\mu$
from below to above  $\mu_{\rm c}$.
Therefore, $\mu$ plays a role analogous to the
inverse temperature in mode-coupling theories.
We then observe (see fig.~\ref{relaxd})
 that after a quench $\rho$ never exceeds $\rho_{\rm c}$, but rather 
approaches it like a power law in time:
\begin{equation}
\rho(\tau) \propto \tau^{-z},
\end{equation}
where $\tau$ is the ``effective time" after the quench (see later)
and the exponent $z$ is approximately equal to $0.3$~.\cite{warning}
Let us remark that $z\phi=1$ within the errors.

The self-correlation function $B(t,t')$ is defined by
\begin{equation}
B(t,t')=\left<|\vec r(t)-\vec r(t')|^2\right>,
\end{equation}
where $\vec r(t)$ and $\vec r(t')$ are the positions of the same
particles at times $t$ and $t'$ respectively.
The average must be defined with some care, since the particles
may leave or enter the system.
We define it by averaging only over the particles which are present
at both times.

From now on we denote the smaller of the two times $t$, $t'$
by $t_{\rm w}$ (the ``waiting time").
Aging corresponds to the fact that $B(t,t_{\rm w})$ does not
depend only on $t-t_{\rm w}$, but rather appears to be
a function of $t/t_{\rm w}$. Similar results have
been obtained for the $p$-spin spherical spin-glass model
\cite{CuKu1}.	
From the figure it is evident that for small values of
$t_{\rm w}$ TTI approximately holds, and aging sets in
only for larger values of $t_{\rm w}$.
In fact, we find here a phenomenon already encountered by Kob and Barrat
\cite{Bako} in Lennard-Jones systems: one has to consider an {\em effective\/}
waiting time that takes into account
the relaxation time of the system before  the quench. 
Hence, one should define an effective waiting time 
$\tau_{\rm w}=t_{\rm w} + \tau_0$, where $\tau_0$ is the relaxation time 
characteristic of the  equilibrium situation \cite{Bako}. 
Figure~\ref{deff} shows that, if one plots $B(t+t_{\rm w},t_{\rm w})$  
vs.\ $t/\tau_{\rm w}$, the curves lie roughly on top of one another. The 
deviations are probably due to finite size effects, e.g.,
to particles that escape from the sample.
Indeed, since the contribution of these particles to
$B(t+t_{\rm w},t_{\rm w})$  
has not been included, the mean squared displacement has a tendency
to be underestimated at long times.
The plot exhibits the waiting-time dependence of the diffusion constant
and is consistent with a simple argument suggesting that diffusion is 
logarithmic in the presence of aging \cite{giorgio}.

\begin{figure}[htbp]
\centering\epsfig{file=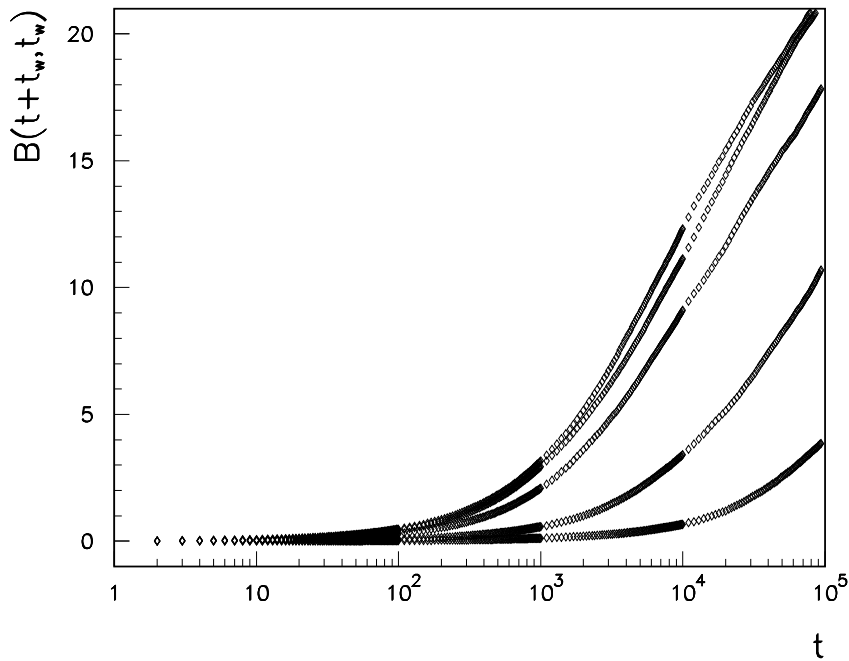,width=0.75\linewidth} 
\caption{Mean square displacement vs.\ time
after a quench to the value $1/\mu= 1/2.2$. 
The waiting times are $t_{\rm w}=10,10^2,10^3,10^4,10^5$ .
The density of initial configuration is $0.75$.
Average over ten samples.}
\protect\label{msd}
\end{figure}

\section{MEAN-FIELD ARGUMENTS}
\begin{figure}[htbp]
\centering\epsfig{file=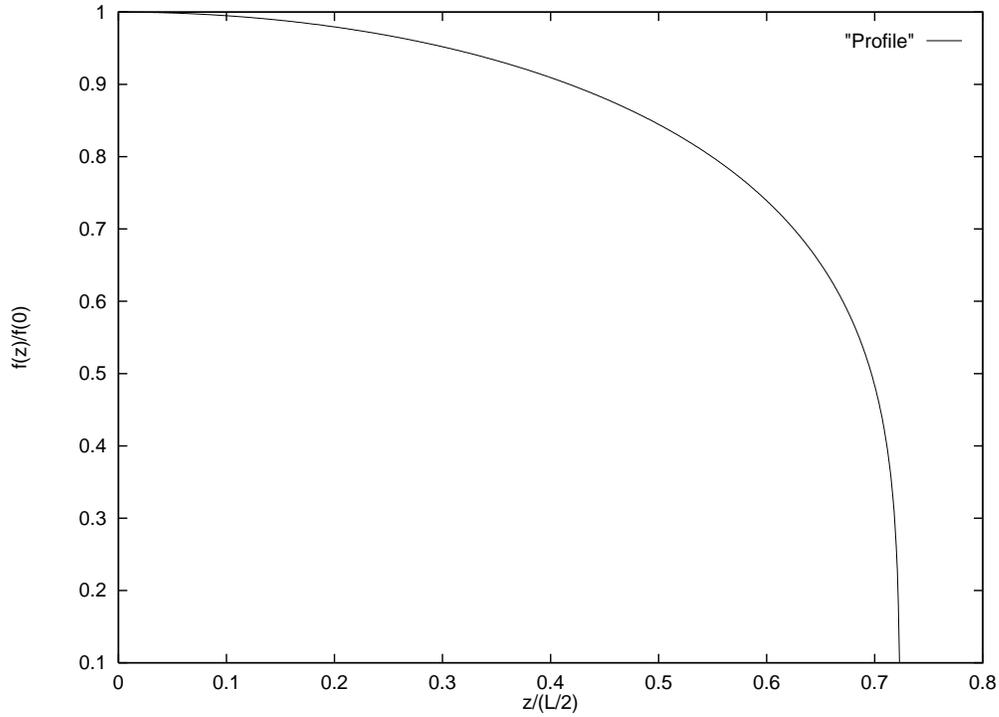,width=0.75\linewidth} 
\caption{Universal density profile $f(z)/f(0)$ vs.\ $z/(L/2)$.
We have chosen the KA value of $\phi=3.1$ .}
\protect\label{profile}
\end{figure}

In this section we show how we can understand the results
reported above by simple mean-filed-like arguments.
Let us assume, as found by KA, that 
$D(\rho) \sim (\rho_{\rm c}-\rho)^{\phi}$,
where, as found by KA, $\phi\simeq 3.1$~.
We consider the sample as a slab, whose
free surfaces at $z=\pm L/2$ are in contact with
the reservoir and therefore rapidly
reach a density equal to the critical one.
Within the sample, the density $\rho(z,t)$ satisfies
the following non-linear diffusion equation:
\begin{eqnarray}
\frac{\partial \rho}{\partial t}
& = & \frac{\partial}{\partial z} \left[D\left(\rho(z,t)\right)
\frac{\partial \rho}{\partial z} \right],
\end{eqnarray}
with the boundary conditions $\rho(L/2)=\rho(-L/2)=\rho_{\rm c}$,
and where $\rho(z,t)<\rho_{\rm c}$ for $-L/2<z<L/2$ and all $t$.
Changing variable $y=\rho-\rho_{\rm c}$ we obtain the equation
\begin{eqnarray}
\frac{\partial y}{\partial t} = \phi y^{\phi-1} 
\left(\frac{\partial y}{\partial z} \right)^2
+y^{\phi} \frac{\partial^2 y}{\partial z^2}.	
\end{eqnarray}
One can look for solutions of the form $y(z,t)=Y(t) f(z)$.
One obtains
\begin{equation}
Y(t) \sim t^{-1/\phi},
\end{equation}
corresponding to a power-law relaxation of the density
with an exponent $z=1/\phi$,
and a differential equation for the density profile $f(z)$:
\begin{equation}
f(z)=\frac{\partial }{\partial z}\left[f^{\phi}(z)f'(z)\right],
\end{equation}
with the boundary conditions
\begin{equation}
f(\pm L/2)=0.
\end{equation}
It is easy to see that this equation allows for solutions of
the form
\begin{equation}
f(z)=f_0 \hat f(z/L),
\end{equation}
where
\begin{equation}
f_0\propto L^{2/(1-\phi)},
\end{equation}
and where the universal profile $\hat f(x)$ satisfies
\begin{equation}
\hat f(x)=\phi \hat f^{\phi-1} (\hat f')^2+\hat f^{\phi}\hat f'',
\end{equation}
where one can choose the boundary conditions $\hat f(0)=f(0)/f_0=1$,
$\hat f'(0)=0.$
The result of a numerical integration of this equation with the
KA value of $\phi=3.1$ is shown in fig.~\ref{profile}.
The solution vanishes for $x=x_0\simeq 0.72$~.
The full solution is therefore $f(z)= f(0)\hat f(2x_0z/L)$.
Unfortunately our simulation data are still too noisy to
allow for a meaningful comparison with this prediction.
 
\begin{figure}[htbp]
\centering\epsfig{file=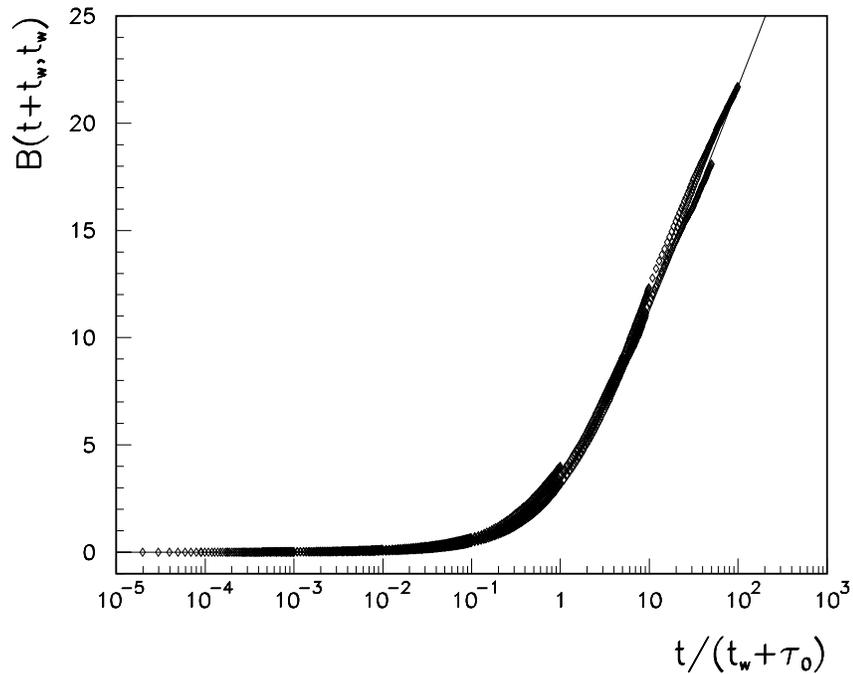,width=0.75\textwidth} 
\caption{Mean squared displacement $B(t+t_{\rm w},t_{\rm w})$ vs.\ 
scaled time $t/\tau_{\rm w}$ (where
$\tau_{\rm w}=t_{\rm w}+\tau_0$)
for $t_{\rm w}=10,10^2,10^3,10^4,10^5$, and 
$\tau_0=10^3$. Quench to the subcritical value $1/\mu=1/2.2$,
starting from density $0.75$. The full line corresponds to
$B(t+t_{\rm w},t_{\rm w})\propto 
(\log (t+t_{\rm w})-\log t_{\rm w})$.
System of size $20^3$, average over ten samples.}
\label{deff}
\end{figure}

Let us now consider the effects of a varying diffusion constant
on the self-correlation function $B(t,t_{\rm w})$. In a simple minded
approach we would obtain
\begin{equation}
B(t,t_{\rm w})=\int_{t_{\rm w}}^t \D t'\, D(t').
\end{equation}
If we assume $D(t)\propto t^{-\zeta}$ we obtain
\begin{equation}
B(t,t_{\rm w})\propto \left(t^{1-\zeta}-t_{\rm w}^{1-\zeta}\right).
\end{equation}
However, in our case, we have $\zeta=z\phi=1$, and a simple integration
leads to 
\begin{equation}
B(t,t_{\rm w})\propto \left(\log t-\log t_{\rm w}\right).
\end{equation}
This is the functional obtained in ref.~\cite{giorgio} from
the hypothesis that $B(t,t_{\rm w})$ depends only on $t/t_{\rm w}$
and from the  ``triangle relation" for $B(t,t')$:
\begin{equation}
B(t,t')=B(t,s)+B(s,t'),\qquad \hbox{for } t'<s<t,
\end{equation}
which stems out of the statistical independence of particle displacements
over nonoverlapping time intervals.
This functional form is borne out by the simulations,
as can be seen by fig.~\ref{deff}.

\section{DISCUSSION}
We find remarkable that this toy model---based as it is upon 
indefendable assumptions---exhibits a behavior so similar to
those of much more complex models. Perhaps the most appealing 
aspect is the fact that aging appears as a consequence of the
slow (power-law) approach to the critical density. This approach
bears some similarity to self-organized criticality: after a sudden
compression, the system endeavors to accommodate as many particles
as it can: but, if the critical density is exceeded, particle
diffusion stops and no more particles can get in. The result
is an ever slower approach to the critical density, and it may be said
that the system ages because it approaches criticality.
Similar considerations in the context of the Bak-Sneppen model
have been made by Boettcher and Paczuski \cite{BP}. In our case they can
be made explicite and quantitative, even within a mean-field-like approach.

Several interesting informations concernig the breakdown of thermodynamical
equilibrium can be obtained from response functions. In our case, one
can consider applying a small constant force $\vec F_{\alpha}$ to particle
$\alpha$ between times $t_{\rm w}$ and $t$. As a consequence, the
particle suffers an extra displacement $\delta\vec r_{\alpha}(t,t_{\rm w})$.
The response function $\chi(t,t_{\rm w})$ is then defined as~\cite{CuLe}
\begin{equation}
\chi(t,t_{\rm w})=\frac{\partial \left<\delta r_i(t,t_{\rm w})\right>}%
{\partial F_i},
\end{equation}
where $i$ is one of the coordinates. The celebrated CK plot \cite{CuKu1}
 of the response
function vs.\ the correlation function should then appear as a broken line,
where the break takes place in correspondence of the value of $B(t,t_{\rm w})$
for which TTI no longer holds. If the ``mean-field"
arguments given above hold, the response of the aging system should
be locked in at what it would exhibit at the critical
density. The slope of the FDT violating line should then
be independent of the quench value of $\mu$. It would be nice to be able to put
this slope in relation with ``universal" properties of the model,
like $\phi$, but we have so far been unable to obtain this relation.

\section*{Acknowledgements} 
It is strange to think that it will not be possible to discuss
this model with Giovannino. We think that he would have liked it,
because it has one of the qualities he appreciated most: coming
close to the heart of a physical problem in simple ways.
We owe much to discussions with Giovannino, and to his playful and
serious ways of making physics. LP would like to apply to Giovannino
and himself the old hebrew saying, in the words of
Lorenzo Da Ponte: {\em E da' discepoli imparai pi\`u
che da tutti}.

Both authors warmly thank Jorge Kurchan for having intitiated
this work and for his continual interest in it.
This work has been done while LP was visiting the Laboratoire de 
Physico-Chimie Th\'eorique of the Ecole Sup\'erieure de Physique et
Chimie Industrielles de la Ville de Paris, supported by a Chaire Joliot.
Part of the work by MS was done during a visit to the same laboratory.
Both authors thank the Laboratoire and its Director, A. Ajdari,
for hospitality and for support. MS also thanks Th.M. Nieuwenhuizen
for support and hospitality at the Van der Waals-Zeeman Laboratorium.


\begin{thebibliography}{99}
\bibitem{warning} The data presented here are the result of more careful
simulations than those quoted in ref.~\cite{KPS}. In particular, the
quench is more ``gentle'' (starts from a value of $\mu$ closer
to $\mu_{\rm c}$), what allows the asymptotic power law behavior to set in
more rapidly. With these simulations, the small discrepancy between 
the values of $\rho_{\rm c}$ with ref.~\cite{KA} has disappeared.
\end{thebibliography}
\end{document}